\documentclass[runningheads]{llncs}

\usepackage{booktabs}
\usepackage{xspace}
\usepackage{graphics}
\usepackage{todonotes}
\usepackage[frozencache=true,cachedir=minted-cache]{minted}
\usepackage[para]{footmisc}
\newcommand{\sys}{\texttt{ir-measures}\xspace}
\newcommand{\treceval}{\texttt{trec\_eval}\xspace}
\newcommand{\cwleval}{\texttt{cwl\_eval}\xspace}
\newcommand{\trectools}{\texttt{trectools}\xspace}
\newcommand{\ndeval}{\texttt{ndeval}\xspace}
\newcommand{\gdeval}{\texttt{gdeval}\xspace}

\usepackage[numbers,sort&compress]{natbib}

\usepackage{url}

\begin{document}
\mainmatter              %
\title{Streamlining Evaluation with \sys}
\author{Sean MacAvaney \and Craig Macdonald
\and Iadh Ounis}
\authorrunning{MacAvaney et al.} %
\institute{University of Glasgow, United Kingdom\\
\email{\{first.last\}@glasgow.ac.uk}}

\maketitle              %

\begin{abstract}
We present \sys, a new tool that makes it convenient to calculate a diverse set of evaluation measures used in information retrieval. Rather than implementing its own measure calculations, \sys provides a common interface to a handful of evaluation tools. The necessary tools are automatically invoked (potentially multiple times) to calculate all the desired metrics, simplifying the evaluation process for the user. The tool also makes it easier for researchers to use recently-proposed measures (such as those from the C/W/L framework) alongside traditional measures, potentially encouraging their adoption.
\end{abstract}
\section{Introduction}

The field of Information Retrieval (IR) is fortunate to have a vibrant and diverse ecosystem of tools and resources. This is particularly true for evaluation tools; there exists a variety of fully-fledged evaluation suites capable of calculating a wide array of measures (e.g., \treceval~\cite{treceval}, \cwleval~\cite{10.1145/3331184.3331398}, \trectools~\cite{palotti2019}, and RankEval~\cite{rankeval-sigir17}) as well as single-purpose scripts that are usually designed for the evaluation of specific tasks or datasets.\footnote{For instance, the MSMARCO MRR evaluation script: \url{https://git.io/JKG1S}} However, none of these tools themselves provide comprehensive coverage of evaluation metrics, so researchers often need to run multiple tools to get all the desired results. Even when a single tool can provide the desired measures, it can sometimes require multiple invocations with different settings to get all desired results (e.g., the TREC Deep Learning passage ranking task~\cite{Craswell2019TrecDl} requires multiple invocations of \treceval with different relevance cutoff thresholds).

In this demonstration, we present a new evaluation tool: \sys.\footnote{Docs: \url{https://ir-measur.es/}, Source: \url{https://github.com/terrierteam/ir_measures}} Unlike prior tools, which provide their own measure implementations, \sys operates as an abstraction over multiple evaluation tools. Researchers are able to simply describe \textit{what} evaluation measures they want in natural syntax (e.g., \texttt{nDCG@20} for nDCG~\cite{Jarvelin:2002:CGE:582415.582418} with a rank cutoff of 20, or \texttt{AP(rel=2)} for Average Precision~\cite{Harman:1992:ESIR} with a binary relevance cutoff of 2), without necessarily needing to concern themselves with which specific tools provide the functionality they are looking for or what settings would give the desired results. By providing both a Python and command line interface and accepting multiple input and output formats, the tool is convenient to use in a variety of environments (e.g., both as a component of larger IR toolkits, or for simply doing \textit{ad hoc} evaluation). By interfacing with existing evaluation toolkits, \sys is more sustainable than efforts that re-implement evaluation measures, especially given the ongoing debate over the suitability of some measures (e.g.,~\cite{Fuhr2018SomeCM,Sakai2020OnFG}) and the proliferation of new measures (e.g.,~\cite{Azzopardi:2021:ECE:3471158.3472239,10.1145/3451161}). An interactive demonstration of the software is available at: \url{https://git.io/JMt6G}.

\section{Background}

Recently, there have been several efforts to resolve incompatibilities for other IR tools. \trectools~\cite{palotti2019} provides Python implementations of numerous IR-related functions including pooling, fusion, and evaluation techniques (including a handful of evaluation measures). PyTerrier~\cite{pyterrier2020ictir} provides a Python interface to a myriad of retrieval, rewriting, learning-to-rank, and neural re-ranking techniques as well as an infrastructure for conducting IR experiments.
CIFF~\cite{Lin2020SupportingIB} defines a common index interchange format, for compatibility between search engines. \texttt{ir\_datasets}~\cite{macavaney:sigir2021-irds} provides a common interface to access and work with document corpora and test collections.
\sys is complementary to efforts like these. In Section~\ref{sec:integratino}, we show that \sys can easily be integrated into other tools, bolstering their evaluation capacity.

\section{\sys}

\sys provides access to over 30 evaluation measures. Table~\ref{tab:measures} provides a summary of the supported measures, which span a variety of categories and applications (e.g., intent-aware measures, set measures, etc.) Measures are referenced by name and a measure-dependent set of parameters (e.g., \texttt{AP(rel=2)} specifies a minimum relevance level and \texttt{nDCG@10} specifies a rank cutoff). We refer the reader to the measure documentation\footnote{\url{https://ir-measur.es/en/latest/measures.html}} for further details.

\begin{table}
\centering
\caption{Measures provided by \sys, along with their providers.}
\label{tab:measures}
\scalebox{0.7}{
\begin{tabular}{|ll|ll|}
\toprule
\bf Measure & \bf Provided by... & \bf Measure & \bf Provided by... \\
\midrule
\texttt{alpha\_nDCG}~\cite{Clarke2008NoveltyAD} & \ndeval & \texttt{NERR}~\cite{Azzopardi:2021:ECE:3471158.3472239} & \cwleval \\
\texttt{(M)AP(@k)}~\cite{Harman:1992:ESIR} & \cwleval, \treceval, \trectools & \texttt{NRBP}~\cite{10.1007/978-3-642-04417-5_17} & \ndeval \\
\texttt{(M)AP\_IA} & \ndeval & \texttt{NumQ}, \texttt{NumRel}, \texttt{NumRet} & \treceval \\
\texttt{BPM}~\cite{Zhang:2017:EWS:3077136.3080841} & \cwleval & \texttt{P(recision)@k}~\cite{rijsbergen:1979:ir} & \treceval, \cwleval, \trectools \\
\texttt{Bpref}~\cite{Buckley2004RetrievalEW} & \treceval, \trectools & \texttt{P\_IA@k} & \ndeval \\
\texttt{Compat}~\cite{10.1145/3451161} & Compatibility script & \texttt{R(ecall)@k} & \treceval \\
\texttt{ERR@k}~\cite{10.1145/1645953.1646033} & \gdeval & \texttt{RBP}~\cite{10.1007/978-3-642-04417-5_17} & \cwleval, \trectools \\
\texttt{ERR\_IA}~\cite{10.1145/1645953.1646033} & \ndeval & \texttt{Rprec}~\cite{Buckley2005RetrievalSE} & \treceval, \trectools \\
\texttt{infAP}~\cite{10.1145/1183614.1183633} & \treceval & \texttt{(M)RR}~\cite{kantor2000trec} & \treceval, \cwleval, \trectools, MSMARCO \\
\texttt{INSQ}~\cite{Moffat:2015:IAM:2838931.2838938}, \texttt{INST}~\cite{Moffat:2012:MMI:2407085.2407092} & \cwleval & \texttt{SDCG@k} & \cwleval \\
\texttt{IPrec@i} & \treceval & \texttt{SetAP}, \texttt{SetF}, \texttt{SetP}, \texttt{SetR} & \treceval \\
\texttt{Judged@k} & OpenNIR script & \texttt{STREC} & \ndeval \\
\texttt{nDCG(@k)}~\cite{Jarvelin:2002:CGE:582415.582418} & \treceval, \gdeval, \trectools & \texttt{Success@k} & \treceval \\
\bottomrule
\end{tabular}
}
\end{table}

\vspace{1em}
\textbf{Providers.} The calculation of measure values themselves are implemented by \textit{providers}. Not all providers are able to calculate all measures (or all parameters of a measure). The current version of \sys includes eight providers:

\treceval~\cite{treceval} is a well-known IR evaluation tool that is used for calculating a variety of measures for TREC tasks. We use Python bindings adapted from the \texttt{pytrec\_eval}~\cite{VanGysel2018pytreceval} package.

\cwleval~\cite{10.1145/3331184.3331398} provides an implementation of a variety of measures that adhere to the C/W/L framework~\cite{10.1145/3052768}, such as \texttt{BPM} and \texttt{RBP}. 

\ndeval\footnote{\url{https://git.io/JKG94}, \url{https://git.io/JKCTo}} enables the calculation of measures that consider multiple possible query intents (i.e., diversity measures), such as \texttt{alpha\_nDCG} and \texttt{ERR\_IA}.

The \trectools~\cite{palotti2019} tookit includes Python implementations of a variety of evaluation measures, including \texttt{AP}, \texttt{Bpref}, and others. \gdeval\footnote{\url{https://git.io/JKCT1}} includes an implementation of \texttt{ERR@k} and an alternative formulation of \texttt{nDCG@k} that places additional weight on high relevance. 

The \textbf{OpenNIR \texttt{Judged@k} script}\footnote{\url{https://git.io/JKG9O}} was adapted from the OpenNIR toolkit~\cite{macavaney:wsdm2020-onir} to calculate \texttt{Judged@k}, a measure of the proportion of top-ranked documents that have relevance assessments.

The \textbf{MSMARCO \texttt{RR@k} script}\footnote{\url{https://git.io/JKG1S}} is an interface to the official evaluation script for the MSMARCO dataset~\cite{Bajaj2016Msmarco2}, with minor adjustments to allow for the configuration of the measure parameters and handling of edge cases. 

The \textbf{Compatibility script\footnote{\url{https://git.io/JKCT5}}} provides \texttt{Compat}~\cite{10.1145/3451161}, a recently-proposed measure that calculates the Rank Biased Overlap of a result set compared to the closest ideal ranking of qrels, which can consider preference judgments.

\vspace{1em}
\noindent \textbf{Interfaces.} \sys can be installed using \texttt{pip install ir-measures}, which provides both a command line interface and a Python package. The command line interface is similar to that of \treceval, accepting a TREC-formatted relevance judgments (qrels) file, a run file, and the desired measures:

\begin{minted}[frame=lines,fontsize=\small]{bash}
$ ir_measures path/to/qrels path/to/run 'nDCG@10 P(rel=2)@5 Judged@10'
nDCG@10  0.6251
P(rel=2)@5  0.6000
Judged@10   0.9486
\end{minted}

Command line arguments allow the user to get results by query, use a particular provider, and control the output format. If the \texttt{ir-datasets}~\cite{macavaney:sigir2021-irds} package is installed, a dataset identifier can be used in place of the qrels path.

The Python API makes it simple to calculate measures from a larger toolkit. A variety of input formats are accepted, including TREC-formatted files, dictionaries, Pandas dataframes, and \texttt{ir-datasets} iterators. The Python API also can provide results by query and can reuse evaluation objects for improved efficiency over multiple runs. Here is a simple example that calculates four measures:

\begin{minted}[frame=lines,fontsize=\small]{python}
> import ir_measures
> from ir_measures import * # import natural measure names
> qrels = ir_measures.read_trec_qrels('path/to/qrels')
> run = ir_measures.read_trec_run('path/to/run')
> ir_measures.calc_aggregate([nDCG@10, P(rel=2)@5, Judged@10], qrels, run)
{nDCG@10: 0.6251, P(rel=2)@5: 0.6000, Judged@10: 0.9486}
\end{minted}

\section{Adoption of \sys}\label{sec:integratino}

\sys is already in use by several tools, demonstrating its utility. It recently replaced \texttt{pytrec\_eval} in PyTerrier~\cite{pyterrier2020ictir}, allowing retrieval pipelines to easily be evaluated on a variety of measures. For instance, the following example show an experiment on the TREC COVID~\cite{Voorhees2020TrecCovid} dataset. \sys allows the evaluation measures to be expressed clearly and concisely, and automatically invokes the necessary tools to compute the desired metrics:

\begin{minted}[frame=lines,fontsize=\small]{python}
import pyterrier as pt
dataset = pt.get_dataset('trec-covid')
pt.Experiment(
  [pt.BatchRetrieve.from_dataset(dataset, "terrier_stemmed")],
  dataset.get_topics("round5"),
  dataset.get_qrels("round5"),
  eval_metrics=[nDCG@10, P(rel=2)@5, Judged@10])
\end{minted}

It is also used by OpenNIR~\cite{macavaney:wsdm2020-onir}, Experimaestro~\cite{10.1145/3397271.3401410}, and DiffIR~\cite{jose:sigir2021-diffir}. The \texttt{ir-datasets}~\cite{macavaney:sigir2021-irds} package uses \sys notation to provide documentation of the official evaluation measures for test collections.

A core design decision of \sys is to limit the required dependencies to the Python Standard Library and the packages for the measure providers (which can be omitted, but will degrade functionality). This should encourage the adoption of the tool by reducing the chance of package incompatibilities.

\section{Conclusion}

We demonstrated the new \sys package, which simplifies the computation of a variety of evaluation measures for IR researchers. We believe that by leveraging a variety of established tools (rather than providing its own implementations), \sys can be a salable and appealing choice for evaluation. We expect that our tool will also encourage the adoption of new evaluation measures, since they can be easily computed alongside long-established measures.

\vspace{1em}

\noindent\textbf{Acknowledgements.} We thank the contributors to the \sys repository. We acknowledge EPSRC grant EP/R018634/1: Closed-Loop Data Science for Complex, Computationally- \& Data-Intensive Analytics.

\bibliographystyle{splncs04nat}
\bibliography{blblio.bib}
\end{document}